\documentclass[
twocolumn,
longbibliography,
showpacs,                     
showkeys,                  
secnumarabic,
amssymb, 
nobibnotes, 
floatfix,
aps, 
pre]{revtex4-1}

\usepackage{graphicx}
\usepackage{latexsym}
\usepackage{amsmath}
\usepackage{verbatim}
\usepackage[justification=justified,singlelinecheck=false]{caption}
\usepackage{caption}
\usepackage{subcaption}
\usepackage{float}
\usepackage{morefloats}
\usepackage{placeins}

\begin{document}

\title{Finite thermal reservoirs and the canonical distribution}

\author{William Griffin}
\author{Michael Matty}

\author{Robert H. Swendsen}
\email[]{swendsen@cmu.edu}
\affiliation{Department of Physics, Carnegie Mellon University, Pittsburgh PA, 15213, USA}

                              
\date{\today}


\begin{abstract}
The 
microcanonical ensemble has long been a starting point for 
the development of thermodynamics from 
 statistical mechanics.
 However, this  approach presents   two problems.
First, it predicts that the entropy is only defined on a discrete set of energies
for finite, quantum systems,
while thermodynamics requires the entropy to be a continuous function 
of the energy.
Second, it fails to  satisfy the stability condition 
($\Delta^2 S / \Delta U^2 < 0$)
for first-order transitions
with both classical and quantum systems.
Swendsen has recently shown that the source of these problems  
lies 
in the microcanonical ensemble itself,
which  contains only energy eigenstates
and excludes  their linear combinations.
To the contrary,
if the system of interest has ever been in thermal contact with 
another system,
it will be described by a probability distribution over many eigenstates
that is equivalent 
to the canonical ensemble
for sufficiently large systems.
Novotny et al. 
have recently  supported 
this picture
by  dynamical numerical calculations 
for a  quantum mechanical model, 
in which they showed the approach to 
a canonical distribution 
for up to 40 quantum spins.
By simplifying the problem to calculate only the equilibrium properties,
we are able to extend the demonstration 
to more than a million particles.
\end{abstract}

\maketitle

\section{Introduction}
\label{introduction}

Since the early work of Boltzmann in classical statistical mechanics\cite{Boltzmann,Boltzmann_translation}
and Planck in quantum statistical mechanics\cite{Planck_1901},
the microcanonical ensemble,
in which the energy is held constant,
has been a fundamental starting point.
However, this use of 
the microcanonical ensemble
 has recently
come under renewed scrutiny
due to two fundamental problems it presents.
\begin{enumerate}

\item
As normally interpreted,
the quantum microcanonical ensemble
is defined by 
the Boltzmann constant times
the logarithm of the degeneracy 
at each discrete energy eigenvalue,
so that
the entropy is not a continuous function of 
the 
energy,
as required by thermodynamics\cite{Planck_1901,Callen,RHS_book,RHS_continuous}.

\item 
In both quantum and classical statistical mechanics, 
for first-order phase transitions 
the microcanonical ensemble
predicts a range of positive second derivatives
of the entropy as a function of energy,
which violates a well-known stability criterion\cite{Callen,RHS_book,RHS_continuous}.

\end{enumerate}

It has been suggested that 
the source of these violations of the thermodynamic postulates
 lies in the neglect of 
quantum states that are linear combinations of 
energy eigenstates\cite{RHS_continuous}.
Such linear combinations are generated 
whenever two macroscopic systems 
exchange energy,
and give a continuous energy spectrum.
%
The canonical ensemble can then
provide an excellent description of 
the statistical properties of a macroscopic system,
even in the absence of  
  contact  with an infinite thermal reservoir.

The traditional derivation of the canonical probability distribution 
involves the expansion of the joint probability distribution 
for the energy of a system of interest 
in thermal contact with a second, much larger
system,
designated a thermal reservoir.
In the limit that the reservoir is 
infinitely larger than the system of interest,
it is then shown that the 
thermal weight
(loosely termed the ``probability'')
of an eigenstate with energy $E_n$
is proportional to 
$\exp( - \beta E_n )$,
where 
$\beta = 1/k_B T$, 
$k_B$ is Boltzmann's constant,
and 
$T$ is the absolute temperature.

We are raising the question of whether the second system
must really be infinitely larger,
or whether it might even be smaller than 
the primary system of interest.
If so,
this would be consistent with normal usage in thermodynamics,
in which a system may be put into thermal equilibrium with
any other macroscopic system regardless of its size 
(Zeroth Law Of Thermodynamics).

Four important papers have recently dealt with 
key aspects of this question\cite{Jin_2013_1,Jin_2013_2,Novotny_decoherence_2015,Novotny_2016_1}.
While we are primarily concerned with the equilibrium features 
of the statistical mechanics of finite systems,
these papers explored the dynamical approach to equilibrium 
under the deterministic time development 
of Newton's equations for classical systems\cite{Jin_2013_1}
and Schr\"odinger's equation for quantum systems\cite{Jin_2013_2,Novotny_decoherence_2015,Novotny_2016_1}.
A consequence of their computations was that 
for both classical and quantum systems 
the projection of 
the joint probability distribution of two finite systems
onto one of the systems
was increasingly well approximated by 
the canonical distribution
as the system sizes increased.
The only limitation of these papers 
was
the practical one 
 that  the solutions of the dynamical equations 
 could only be carried out for relatively small systems.

In this paper,
we take a simpler approach to a more limited problem 
than was treated in 
Refs.~\cite{Jin_2013_1,Jin_2013_2,Novotny_decoherence_2015,Novotny_2016_1}.
We restrict ourselves to the equilibrium statistical mechanics
of finite systems,
so we ignore the time development of the 
microscopic states.
This means that we only need the information 
contained in 
 the degeneracies of the quantum energy levels, $ \omega ( E_n )$,
and their equilibrium occupation.
We will show that under very general conditions,
the distribution of energies is very nearly canonical.
This in turn has the consequence that the thermodynamic entropy 
as a function of energy 
is given by 
the canonical entropy, 
and not the microcanonical.

In the next section,
we give the basic definitions for analyzing 
two interacting quantum systems,
including the joint probability distribution in equilbirum,
and the canonical distribution as an approximation 
to that distribution.
In Section \ref{section_Massieu},
we recall the definition of the Massieu function,
which is a version of the Helmholtz free energy
that is applicable to both positive and negative temperatures.
Section \ref{section_SHO}
then defines 
$\delta$,
which 
was introduced in 
Refs.~\cite{Jin_2013_1,Jin_2013_2,Novotny_decoherence_2015,Novotny_2016_1}
 a measure of the difference between 
the joint probability distribution
and the canonical approximation.
A system of simple harmonic oscillators 
is used to illustrate how a finite reservoir 
can generate an excellent approximation to the canonical distribution --
even if it is smaller than the system of interest.
In Section \ref{section_Ising} 
we discuss the behavior of the two-dimensional Ising model, both at and away from the 
second-order phase transition.
In Section \ref{section_first_order},
we discuss a corresponding analysis of the two-dimensional, twelve-state Potts model,
 both at and away from the 
first-order phase transition.

\section{Joint probability distribution of two finite quantum systems}
\label{section: quantum joint probability}

Although thermodynamics and statistical mechanics 
also make predictions for other macroscopic variables,
we will limit consideration to energy distributions
to represent the general situation.
Similarly,
although a general treatment would include 
the exchange of energy between an arbitrary number 
$M  \ge 2$ subsystems, 
restrict ourselves to 
 $M=2$ subsystems,
since the generalization to more subsystems is obvious.
Label the systems with $S$ (for the system of interest)
 and $E$ (for the environment).
The joint Hamiltonian is then denoted as
\begin{equation}
H_{\textrm{total}}
=
H_S  +  H_E +   L_{S,E}  ,
\end{equation}
where 
$H_S$ is the Hamiltonian of the system of interest,
$H_E$ is the Hamiltonian of the environment,
and
$L_{S,E}$ 
denotes the interactions between 
particles in different subsystems.
When the systems are separated,
each system has its own energy spectrum.
For simplicity,
we will assume that the separation between energy eigenvalues
is a multiple of $\epsilon$ in both the system and the environment.
Let us assume that the system and the environment together are separated from the rest of the universe,
so that the total energy is a constant $E_T$.
When the system and the environment are separated,
let the density of states  of the system be $\omega_S(E_n)$
(the degeneracy of energy level $E_n$),
and 
$\omega_E(E_T-E_n)$ for the environment,
where $n$ denotes an eigenstate of the system.

It is normally the case that the energies associated with
$L_{S,E}$ 
 are negligible  in comparison with the other energies in the problem.
 Nevertheless, they allow the transfer of  energy between the two systems, 
 by changing the energy spectrum from two separate spectra 
 to a single spectrum\cite{RHS_continuous}.

 After the two systems have been brought together, equilibrated, and then separated, the energy distribution in each system is
 not a delta function.
 Each combination of states will be equally likely, so that the probability of two energy levels with total energy  
 $E_T$
  will have a probability proportional to the product of the degeneracies
  $\omega_S ( E_n ) \omega_E ( E_T  - E_n )$.
  Normalizing, this becomes the probability for energy level $n$ in the original system.
 \begin{equation}\label{product probability}
P_S (E_n, E_T)
=
\frac{\omega_S ( E_n ) \omega_E ( E_T  - E_n )}
{  \sum_{E_n} \omega_S ( E_n ) \omega_E ( E_T  - E_n )}
\end{equation}
 The equilibrium value of the energy in $S$
  is given by
  \begin{equation}\label{product equilibrium value}
\langle  E_n \rangle_S
=
\sum_{E_n}
E_n
P_S (E_n, E_T)
\end{equation}

  This should be compared to the probability for the energy level $n$
   of the system in contact with the thermal reservoir. 
   This probability can be derived from
   Eq.~(\ref{product probability})
  by expanding the  logarithm of $\omega_E$
   about its equilibrium value.
\begin{eqnarray}
\ln P_S (E_n, E_T)
&=
\ln \omega_S ( E_n )
+
\ln \omega_E ( E_T  - E_n )    \nonumber  \\
&
 - \ln   \left[
  \sum_{E_n} \omega_S ( E_n ) \omega_E ( E_T  - E_n )  \right]
\end{eqnarray}

 \begin{eqnarray}
\ln P_S (E_n, E_T)
&=
\ln \omega_S ( E_n )
+
\ln \omega_E \left( E_T  - \langle  E_n \rangle_S \right)    \nonumber \\
&
-
 (E_n - \langle  E_n \rangle_S ) 
 \frac{ \partial }{ \partial E_T }
\ln \left[ \omega_E ( E_T  - \langle  E_n \rangle_S ) \right]  \nonumber \\
&
 +  (\textrm{higher order terms})      \nonumber \\
&
 - \ln  \left[
  \sum_{E_n} \omega_S ( E_n ) \omega_E ( E_T  - E_n ) \right]
\end{eqnarray}
 The factor 
 $ \beta = \frac{ \partial }{ \partial E_T }
\ln \left[  \omega_E ( E_T  - \langle  E_n \rangle_S ) \right]$
  can be interpreted as the inverse temperature.

 Collecting terms, and dropping the higher order terms, we have
 \begin{equation}
\ln P_S (E_n, E_T)
\approx
\ln P_{S,C}(E_n, \beta)  ,
\end{equation}
where
\begin{equation}
\ln P_{S,C}(E_n, \beta)
=
\ln \omega_S ( E_n )
- \beta (E_n - \langle  E_n \rangle_S ) 
-\ln X     ,
\end{equation}
or 
 \begin{equation}
 P_{S,C} (E_n,\beta)
=
\frac{1}{X}
 \omega_S ( E_n )
\exp [- \beta (E_n - \langle  E_n \rangle_S ) ]     .
\end{equation}
This is just the canonical distribution 
\begin{equation}
 P_{S,C} (E_n,\beta)
=
\frac{1}{Z}
 \omega_S ( E_n )
\exp [- \beta E_n ]
\end{equation}
with different parameters
because it is expanded about 
$\langle  E_n \rangle_S$
instead of zero.
The constants are related by
\begin{equation}
\frac{1}{Z}
=
\frac{1}{X}
\exp [ \beta \langle  E_n \rangle_S ) ]
\end{equation}

It is clear that 
$ P_S (E_n, E_T)$
and 
$P_{S,C} (E_n, E_T)$
differ,
but the question is whether this difference is large enough to matter.
The question is of fundamental importance.
If they differ measurably for macroscopic systems,
it is not sufficient to specify the temperature 
of system $E$,
but the system size must also be known.
If the difference is not measurable,
the assumptions of thermodynamics
are valid.
Furthermore,
the entropy as a function of energy
(the fundamental relation) 
must be given by the canonical entropy,
which we now derive.

\section{Massieu functions and the canonical entropy}
\label{section_Massieu}

It is well known that the canonical partition function 
is related to the Helmholtz free energy,
 $F=U-TS$,
by the equation
\begin{equation}
\ln Z ( \beta, V,N ) = - \beta F(T,V,N)  
\end{equation}
%
The Helmholtz free energy is,
of course,
the Legendre transform of the 
fundamental relation
$U=U(S,V,N)$
with respect to temperature,
which we will denote as 
$F(T,V,N) = U[T]$,
indicating the Legendre transform 
by the square brackets
around the new variable $T$\cite{Callen,RHS_book}.
For  generality,
it is better the use a Massieu function,
which is the Legendre transform of the entropy\cite{Callen,RHS_book}.
The reason is that if 
$S=S(U,V,N)$
is not monotonic in $U$,
the function cannot be inverted
to find 
$U=U(S,V,N)$.
It will be particularly useful to define a dimensionless entropy,
$\tilde{S}=S/k_B$,
in forming Massieu functions.

From the differential form of the fundamental relation
for $dS$,
we can see that
\begin{equation}
d \tilde{S} = \beta \, dU + \beta P dV - \beta \mu \, dN,
\end{equation}
where 
$P$ is the pressure,  $V$ is the volume,
$\mu$ is the chemical potential, and $N$ is the number of particles.
The inverse temperature $\beta$ is found from the usual equation,
which can be written as
\begin{equation}\label{S[b]=S-bU}
\beta
=
\left(
\frac{ \partial \tilde{S} }{ \partial U }   
\right)_{V,N}  .
\end{equation}
The Legendre transform (Massieu function) is given by
\begin{equation}\label{s[b]= - b F}
\tilde{S}[\beta] = \tilde{S} - \beta U = - \beta \left( U - TS \right) = - \beta F ,
\end{equation}
so that 
\begin{equation}\label{St b = ln Z}
\tilde{S}[\beta] =\ln Z ( \beta, V,N )   .        
\end{equation}
The differential form 
of the Massieu function $\tilde{S}[\beta]$
is then
\begin{equation}
d \tilde{S}[\beta] = - U d \beta + \beta P dV - \beta \mu dN .
\end{equation}
This immediately gives us
\begin{equation}\label{d tilde S / dbeta = - U}
\left(
\frac{ \partial \tilde{S}[ \beta ] }{ \partial \beta  }   
\right)_{V,N}  =
- U   
=
-
\left(
\frac{ \partial  ( \beta F )  }{ \partial \beta  }   
\right)_{V,N} ,
\end{equation}
where the last equality is a well-known thermodynamic identity\cite{Callen,RHS_book}.

%


To carry out the inverse Legendre transform of 
$\tilde{S}[\beta]$
to find 
$S(U)$,
use 
Eq.~(\ref{d tilde S / dbeta = - U})
to find 
$U=U(\beta)$.
Since $U$ is a monotonic function of 
$\beta$, 
even for a non-monotonic density of states,
we can invert this equation to obtain 
$\beta=\beta(U)$.
From 
Eq.~(\ref{S[b]=S-bU}),
we can find
\begin{equation}\label{St = St b + b U}
\tilde{S} = \tilde{S}[\beta] +\beta(U) U .
\end{equation}
Finally, the entropy with the usual dimensions is given by 
\begin{equation}
S
=
k_B \tilde{S}  .
\end{equation}

%

\section{Comparison of canonical and joint distributions 
for simple harmonic oscillators}
\label{section_SHO}

A set of simple harmonic oscillators (SHO)
provides a test for these ideas\cite{RHS_continuous}.   
The energy spectrum is 
\begin{equation}\label{SHO H}
E = \hbar \omega \sum_k^N  \left( n_k  + \frac{1}{2} \right)  ,
\end{equation}
where $\hbar$ is Planck's constant,
and $\omega$ is the angular frequency,
and $n_k$ takes on the values of all nonnegative integers.
%
The canonical entropy is 
\begin{equation}
S_{\textrm{SHO}}
=
N k_B
\left[
- x \ln x
+ ( 1 + x ) \ln (1+x) ,
\right]
\end{equation}
where
\begin{equation}
x
=
\frac{
U_{\textrm{SHO}} -\frac{1}{2} 
\hbar \omega N
}{
N \hbar \omega 
}   .
\end{equation}

To provide a numerical measure for the 
agreement between the canonical entropy 
and the joint distribution,
we follow Novotny et al.\cite{Novotny_decoherence_2015,Novotny_2016_1}
in defining $\delta$.
\begin{equation}\label{delta_1}
\delta
=
\sqrt{
\sum_{E_n}
\left(
P_S(E_n, E_T)
-
P_{S,C}(E_n, \beta )
\right)^2
}
\end{equation}
where  
\begin{equation}
P_{S,C}(E_n, \beta )
=
D_S(E_n)
e^{-\beta E_n} / Z    ,
\end{equation}
with $D_S(E_n)$ to denote the density of states of the system $S$
(to avoid confusion between $\omega$ and $\omega_S(E_n)$ ),
and
\begin{equation}
Z = \sum_{E'_n} 
D_S(E'_n)
e^{-\beta E'_n}      .
\end{equation}

The value of
$\delta$ is zero when the two distributions are identical,
which is the case for 
$N \rightarrow \infty$.
For finite systems,
it indicates how much the two distributions
differ.
Therefore,
a small value of delta means that the canonical
distribution provides a good description of the system.

We investigated the probability distributions for 
two systems of simple harmonic oscillators 
representing the system (S) and the environment (E).
We used sizes ranging from 
$10$ to $10^6$.
The energy was fixed at $1.5\epsilon$  per oscillator 
for  $N_S$ and $N_E$
between $10$ and $10^6$.
Table~\ref{ table:_SHO_1} gives the values of 
$\delta$.

\begin{table}[htb]
\caption{The values of $\delta$  
for a system of $N_S$ simple harmonic oscillators
given in 
Eq.~(\ref{SHO H})
from computations 
with  environment composed of  $N_E$
simple harmonic oscillators.
The values of the system size, $N_S$, 
are given as headings of the columns. 
The average energy is $1.5$,
and the temperature is  $1.44$.
}
\begin{center}
\begin{tabular}{c||c|c|c|c|c|c}
$N_E$   & 	 $ 10	 $ &  $ 10^2 $ 	&  $ 10^3 $ 	&  $ 10^4 $ 	&  $ 10^5 $ 	&  $ 10^6 $   \\
\hline
 $ 10 $ 	&   8.61E-2   	&    2.19E-1    
	        &    2.68E-1 	           &    2.61E-1    
	        &    2.58E-1    	&    2.57E-1     \\
 $ 10^2 $ 	&    1.05E-2 	   &     4.61E-2    	
            &         1.03E-1 	   &    1.33E-1 	   
               &      1.43E-1    	&    1.42E-1       \\
 $ 10^3 $ 	&    1.08E-3    	 &    5.90E-3 	   
                &    2.57E-2      & 	   5.68E-2 	   
                &    7.24E-2 	   &    7.79E-2      \\
 $ 10^4 $ 	&    1.08E-4      &    6.09E-4  	
            &        3.32E-3   &    1.45E-2 	 
               &     3.19E-2   &   4.06E-2     \\
  $ 10^5 $ 	&    1.08E-5   &    6.11E-5 	 
            &        3.43E-4     &    1.86E-3 	 
               &      8.13E-3 	 &    1.79E-2    \\
 $ 10^6 $ 	&   1.08E-6   &   6.11E-6 	 
            &     3.44E-5 	 &   1.93E-4 	   
               &    1.05E-4   &     4.57E-3  
\end{tabular}
\end{center}
\label{ table:_SHO_1}
\end{table}

\begin{figure}  [ht]
 \includegraphics[width=\columnwidth]{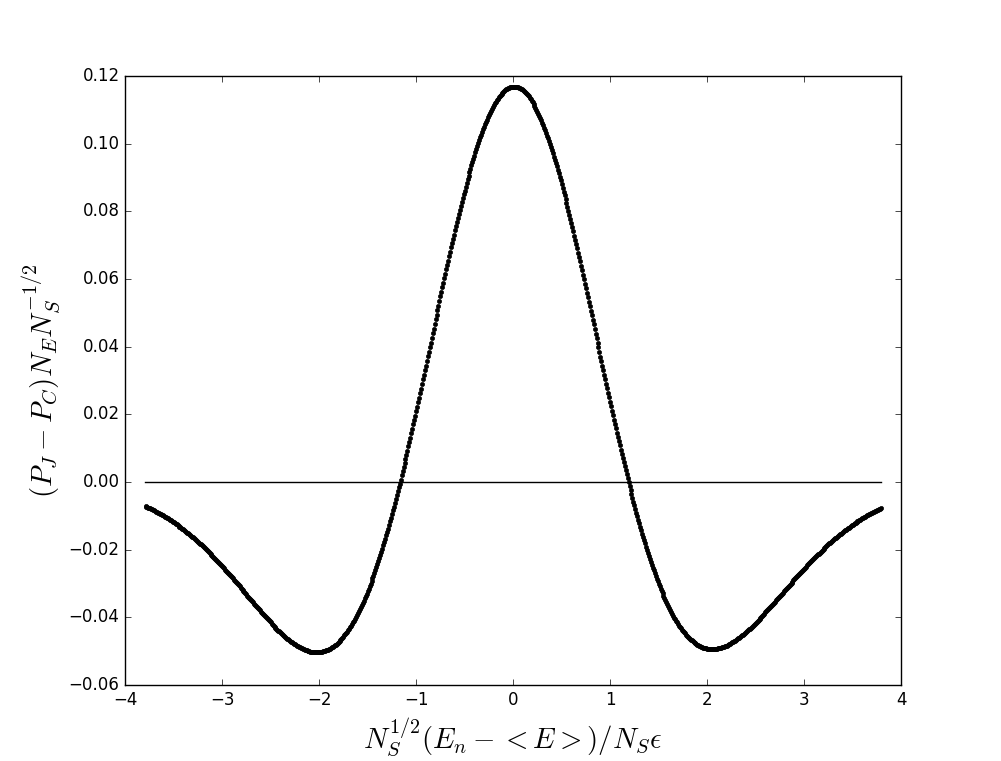}
  \caption{
  The scaled difference in the probabilities 
  $(P_J-P_C)N_E/N_S^{1/4}$
  is plotted against the 
  scaled system energies
  $ N_S^{1/2}(E_n - \langle E \rangle ) / N_S\epsilon $.
  The system contained $N_S=10^4$ SHO's,
  while the environment contained $N_S=10^4$ SHO's
  ($R=N_S/N_E=1$).
The asymptotic behavior is shown,
and the plot will not change for larger systems.
   The effective inverse temperature is $1.0$,
  and the average energy is $1.44$.
  }
  \label{SHO_diff_10000_10000}
\end{figure}

To understand the convergence of $\delta$,
first consider the dependence on the size of the environment, $N_E$.
Table~\ref{ table:_SHO_1} shows clearly that $\delta \rightarrow 1/N_E$,
which is due to the quadratic dependence 
of the difference 
on the energy
between $P_S(E_n, E_T)$ 
and 
$P_{S,C}(E_n, \beta )$.
The  size of the system, $N_S$,
enters in two places.
The width of the distribution is proportional to 
$\sqrt{N_S}$,
which is a standard result for the width of a peak
in statistical mechanics.
This suggests scaling $\left( E_n-\langle E \rangle \right)/N_S$
by 
$1/\sqrt{N_S}$.
To maintain the normalization of the probabilities,
we multiply them by $\sqrt{N_S}$.
The scaling behavior of the difference in the probabilities
is illustrated in
Fig.~\ref{SHO_diff_10000_10000}.

The scaling dependence of $\delta$ is then given by
\begin{equation}\label{delta_scaled_1}
\delta  =  \tilde{\delta} (N_S/N_E)  N_S^{3/4}  N_E^{-1}  
\end{equation}
where $ \tilde{\delta}  (N_S/N_E)$  is a function of  the ratio
$R=N_S/N_E$ for systems with either $N_S$ or $N_E$
larger than about $10^4$.

\begin{table}[hb]
\caption{The values of $\tilde{\delta}$  
(the rescaled values of $\delta$)
for a system of $N_S$ simple harmonic oscillators
given in 
Eq.~(\ref{SHO H})
from computations 
with  environment composed of  $N_E$
simple harmonic oscillators.
The values of the system size, $N_S$, 
are given as headings of the columns. 
The average energy is $1.5$,
and the temperature is  $1.44$.
}
\begin{center}
\begin{tabular}{c||c|c|c|c|c|c}
$N_E$   & 	 $ 10	 $ &  $ 10^2 $ 	&  $ 10^3 $ 	&  $ 10^4 $ 	&  $ 10^5 $ 	&  $ 10^6 $   \\
\hline
 $ 10 $ 	& 0.15314 & 0.06917 & 0.01507 & 0.00261 & 0.00046 & 
 0.00008 \\
 $ 10^2 $ 	&  0.18610 & 0.14582 & 0.05795 & 0.01330 & 0.00255 & 0.00045 \\
 $ 10^3 $ 	& 0.19155 & 0.18645 & 0.14472 & 0.05679 & 0.01287 & 0.00246 \\
 $ 10^4 $ 	& 0.19215 & 0.19255 & 0.18649 & 0.14461 & 0.05666 & 0.01282 \\
 $ 10^5 $ 	& 0.19221 & 0.19319 & 0.19266 & 0.18649 & 0.14459 & 0.05665 \\
 $ 10^6 $ 	& 0.19221 & 0.19325 & 0.19331 & 0.19267 & 0.18649 & 0.14459
\end{tabular}
\end{center}
\label{ table:_SHO_1_delta_tilde}
\end{table}

As an illustration,
consider the following cases.
\begin{enumerate}
\item If the environment is much larger than the system
($N_S<<N_E$), then $ \tilde{\delta} (R)$ becomes 
a constant  
as $R \rightarrow 0$,
as seen in  the bottom left side of 
Table~\ref{ table:_SHO_1_delta_tilde}.

\item  For a fixed  ratio 
$R=N_S/N_E$,
$\delta \rightarrow N^{-1/4}$,
as the sizes of the two systems increase
(where $N$ is either $N_S$ or $N_E$).
This can be seen by looking at any diagonal in 
Table~\ref{ table:_SHO_1_delta_tilde},
where
the values of $\tilde{\delta}$ go to a constant.
that depends on $R$.
This decrease is rather  slow,
but the deviation from canonical behavior 
might still be hard to measure for a macroscopic system,
as seen in Table~\ref{table_fixed_ratio}.

\begin{table}[htb]
\caption{The values of $\delta$ for 
$N_S=10^{20}$
as a function of the ratio 
$R=N_S/N_E$.
}
\begin{center}
\begin{tabular}{c|c}
$R$  & 	  $\delta$	   \\
\hline
0.001  &  1.93E-9   \\
0.01  &  1.93E-8   \\
0.1  &  1.86E-7    \\
1   &  1.45E-6   \\
10  &   5.66E-6  \\
100   &  1.28E-5 \\
1000   &  2.46E-5 
\end{tabular}
\end{center}
\label{table_fixed_ratio}
\end{table}

\item  For the case that $N_E$ is large,
$\tilde{\delta}$
is independent of $N_S$,
and 
$\delta$
goes as $N_S^{3/4}$,
as given by Eq.~(\ref{delta_scaled_1}),
and shown in Tables~\ref{ table:_SHO_1}
and \ref{ table:_SHO_1_delta_tilde}.

\end{enumerate}

The numerical calculations have confirmed  the predictions of  Eq.~(\ref{delta_scaled_1}).


\section{Comparison of canonical and joint distributions 
for 
the Ising model
(second-order phase transitions)}
\label{section_Ising}

The exchange of energy between an Ising model and 
a system of simple harmonic oscillators
is different at the second-order phase transition
and away from it.
We will examine both cases.

\subsection{Away from the second-order phase transition}

We did calculations for an inverse temperature of  $\beta=0.5$,
which is well away from the critical value of 
$\beta_c=0.88137$.
The energy was set as close to the average energy 
as possible,
given the discrete spectrum.
%

  
To obtain a quantitative measure of the agreement,
we compared a variety of sizes for both the system and the environment.
The canonical distribution 
was used to obtain the best fit for the temperature.
The deviation of the temperature from the desired value
was small for all systems,
and got dramatically smaller as the systems got larger.
As can be seen from 
Tables~\ref{ table:_Ising_b5_1_delta} 
and \ref{ table:_Ising_b5_1_delta_tilde},
the data shows the same scaling given  
in
Eq.~(\ref{delta_scaled_1}).


\begin{table}[htb]
\caption{The values of $\delta$  
for a system of $L \times L$ Ising models
from computations 
with  environment composed of  $N_E$
simple harmonic oscillators.
The values of the system size, $N_S = L^2$, 
are given as headings of the columns. 
The average energy is  $-1.279$,
and the inverse temperature is  $\beta=0.5$.}
\begin{center}
\begin{tabular}{c|c|c|c}
$N_E$ &   $8 \times 8$  & $16 \times 16$  & $32 \times 32$  \\
\hline
$2^6$  &	 2.05E-02	&5.04E-02 &	9.69E-02  \\
$2^8$  	&    5.48E-03 &	1.46E-02&	3.56E-02\\
$2^{10}$   & 	1.40E-03 &	3.84E-03 &	1.04E-02\\
$2^{12}$   & 	 3.51E-04 &	9.72E-04 &	2.72E-03\\
$2^{14}$   & 	 8.79E-05 &	2.44E-04 &	6.88E-04\\
$2^{16}$   &  2.20E-05 &	6.10E-05 &	1.72E-04
\end{tabular}
\end{center}
\label{ table:_Ising_b5_1_delta}
\end{table}


\begin{table}[htb]
\caption{The values of $\tilde{\delta}$  
(the rescaled values of $\delta$)
for a system of $L \times L$ Ising models
from computations 
with  environment composed of  $N_E$
simple harmonic oscillators.
The values of the system size, $N_S = L^2$, 
are given as headings of the columns. 
The average energy is $-1.279$,
and the inverse temperature is  $\beta=0.5$.
}
\begin{center}
\begin{tabular}{c|c|c|c}
$N_E$ &   $8 \times 8$  & $16 \times 16$  & $32 \times 32$  \\
\hline
$2^6$  &	 0.05804&	0.05044 &	0.03424  \\
$2^8$  	&    0.06197 &	0.05852 &	0.05033  \\
$2^{10}$   & 	0.06322 &	0.06139 &	0.05866  \\
$2^{12}$   & 	 0.06359 &	0.06221 &	0.06146  \\
$2^{14}$   & 	0.06368 &	0.06244 &	0.06223  \\
$2^{16}$   &   0.06369 &	0.06249 &	0.06227 
\end{tabular}
\end{center}
\label{ table:_Ising_b5_1_delta_tilde}
\end{table}

%
\subsection{At  the second-order phase transition}

\begin{figure}[htb]
   \begin{subfigure} [htb] {0.3\textwidth}
 \includegraphics[width=\columnwidth]{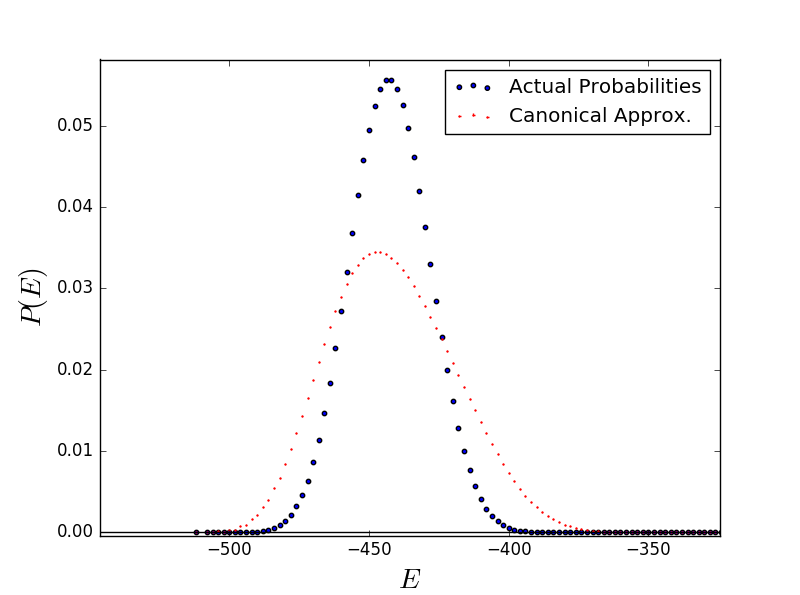}
  \caption{
 $2^{8}$ SHO's as environment.
      }
  \label{256ising_256osc}
  \end{subfigure}
 \begin{subfigure} [htb] {0.3\textwidth}
 \includegraphics[width=\columnwidth]{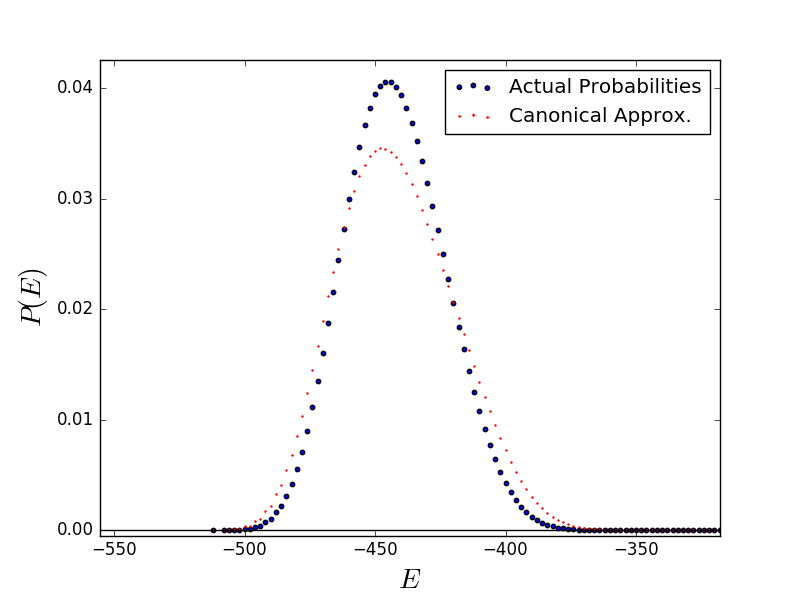}
  \caption{
$2^{10}$ SHO's as environment.
      }
  \label{256ising_1024osc}
  \end{subfigure}
 \begin{subfigure} [htb] {0.3\textwidth}
 \includegraphics[width=\columnwidth]{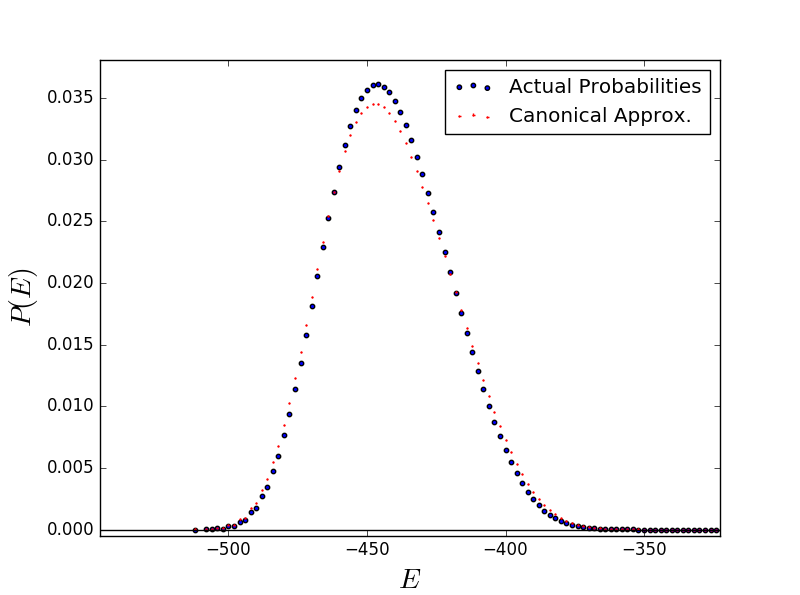}
  \caption{
$2^{12}$ SHO's as environment..
        }
  \label{256ising_4096osc}
\end{subfigure}
\caption{    Comparison of the canonical distribution (black dots)
  with the true joint distribution (red dots)  for a
Ising model 
  on a $16 \times 16$ lattice 
  at the transition energy
  interacting with a system of simple harmonic oscillators.
  The system is at the critical  inverse temperature.}
\label{Ising_256_SHO}
\end{figure}

The width of the peak in the energy is largest at the second-order phase transition
due the connection between the variance of the energy and the specific heat.
\begin{equation}
c_S = C_S/N_S=
\frac{1}{N_S k_B T^2} 
\left\langle  \left(  E - \langle  E \rangle \right)^2 \right\rangle
\end{equation}
The increased width of the energy peak 
requires a larger environment to show a good approximation
to the canonical ensemble.
The improvement of the fit with increasing size is shown in
Fig.~\ref{Ising_256_SHO}.

 
%
The increase of the width of the critical energy peak also
has the effect of increasing the value of $\delta$,
which is shown in 
Table~\ref{ table:_Ising_critical_1_delta}.
The values should be compared to 
Table~\ref{ table:_Ising_b5_1_delta}
for $\beta \ne \beta_c$,
which are all smaller.


\begin{table}[htb]
\caption{The values of $\delta$  
for a system of $L \times L$ Ising models
from computations 
with  environment composed of  $N_E$
simple harmonic oscillators.
The values of the system size, $N_S = L^2$, 
are given as headings of the columns. 
The  temperature is  $\beta_c=0.88137$.
}
\begin{center}
\begin{tabular}{c|c|c|c}
$N_E$ &   $8 \times 8$  & $16 \times 16$  & $32 \times 32$  \\
\hline
$2^6$  &	1.00E-01 &	1.69E-01 &	2.14E-01    \\
$2^8$  	&   3.09E-02 &	7.46E-02 &	1.24E-01    \\
$2^{10}$   & 	8.18E-03 &	2.42E-02 &	5.73E-02    \\
$2^{12}$   & 	 2.08E-03 &	6.58E-03 &	1.95E-02    \\
$2^{14}$   & 	 5.21E-04 &	1.68E-03 &	5.42E-03    \\
$2^{16}$   &  1.30E-04 &	4.22E-04 &	1.39E-03
\end{tabular}
\end{center}
\label{ table:_Ising_critical_1_delta}
\end{table}

We should be able to rescale $\delta$ 
for the Ising critical point
to demonstrate the effect of the specific heat.
Unfortunately,
we have not been able to do this,
probably because the systems are not large enough
to show the asymptotic behavior.


\section{Comparison of canonical and joint distributions 
for first-order transitions
($12$-state Potts model)}
\label{section_first_order}

The twelve-state Potts model in two dimensions 
presents an interesting test of finite-canonical distributions.
Away from the first-order transition,
it shows the same 
finite-canonical behavior distribution
as the system of simple harmonic oscillators
seen in the previous section.
At the first-order transition,
the behavior is different.

In each case, the energy 
was chosen to correspond as closely as possible 
 to the temperature being investigated
($\beta=1.0$ and $\beta_c=1.496068$).
The value of the inverse temperature 
used for the comparison with the canonical distribution
was then optimized to reduce the value of $\delta$.
This temperature correction was about $1\%$
for the smallest example ($8 \times 8$ Potts lattice and $64$ SHOs),
but went down to $7.6 \times 10^{-6}$
for a $32 \times 32$ lattice and $N_E=2^{18}$.

We first demonstrate the behavior away from the transition.

\subsection{Away from the first-order transition}

Values of $\delta$ were calculated for an inverse temperature of
$\beta=1.0$,
far from the first-order transition at 
$\beta_c=1.49607$.
The scaled values for 
 $\tilde{\delta}$
 are  given in  Table~\ref{table:_Potts_optimized_beta_1.0_delta_tilde}.
The patterns is the same as seen in
Table~\ref{ table:_SHO_1_delta_tilde},
although the values of   $\tilde{\delta}$
are somewhat smaller.
The asymptotic scaling of $\delta$
is confirmed.

\begin{table}[htb]
\caption{The values of $\tilde{\delta}$  
(the rescaled values of $\delta$)
for a two-dimensional, $12$-state Potts model
at $\beta \approx 1.0$ (away from the first-order transition).
The size of the environment
(a set of simple harmonic oscillators)
is shown in the left column.
Values of $\delta$ 
for the three sizes of the Potts model investigated
($8 \times 8$, $16 \times 16$, and $32 \times 32$) 
are given in the next three columns.
}
\begin{center}
\begin{tabular}{r|r|r|r}
$N_E$ &   $8 \times 8$  & $16 \times 16$  & $32 \times 32$  \\
\hline
$2^6$  & 0.10519	 & 0.08274	 & 0.04579 \\
$2^8$  	 & 0.11993	 & 0.11032	 & 0.08200\\
$2^{10}$  & 0.12432	 & 0.12291	 & 0.11048  \\
$2^{12}$     & 0.12556	 & 0.12691	 & 0.12289 \\
$2^{14}$    & 0.12592	 & 0.12803	 & 0.12671 \\
$2^{16}$     & 0.12599	 & 0.13292	 & 0.12769  \\
$2^{18}$  	 & 0.12605	 & 0.12837	 & 0.12794
\end{tabular}
\end{center}
\label{table:_Potts_optimized_beta_1.0_delta_tilde}
\end{table}

\subsection{At the first-order transition}

\begin{figure}[htb]
\begin{center}
   \begin{subfigure} [htb] {0.3\textwidth}
    \includegraphics[width=\columnwidth]{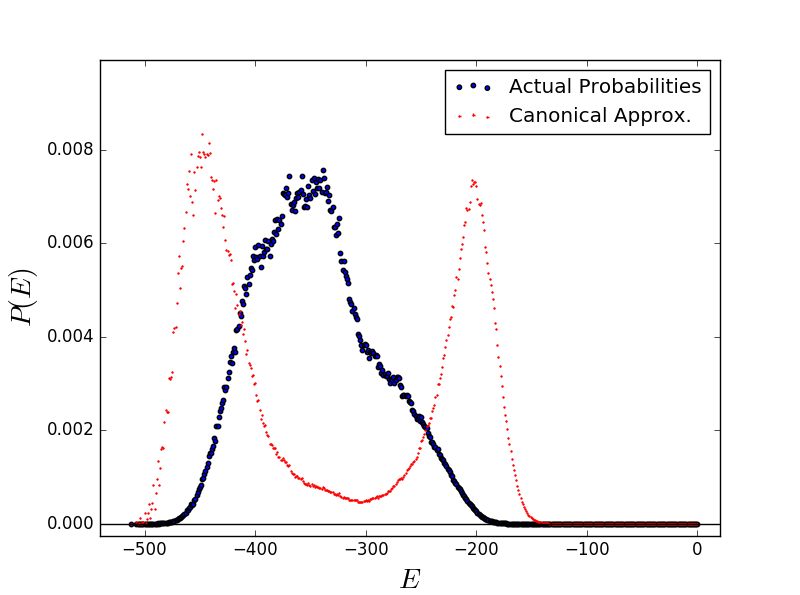}
  \caption{
$2^{12}$ simple harmonic oscillators.
      }
  \label{Potts_256_SHO_4096}
  \end{subfigure}
 \begin{subfigure} [htb] {0.3\textwidth}
 \includegraphics[width=\columnwidth]{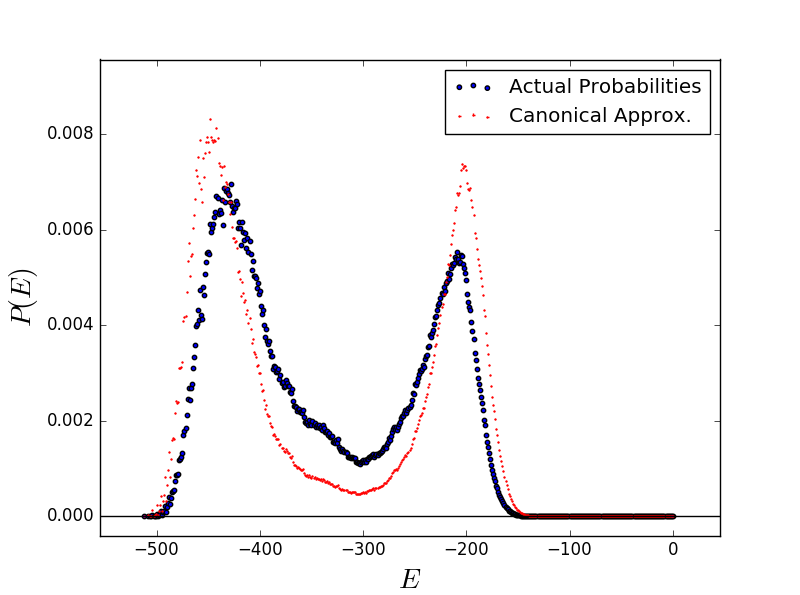}
  \caption{
$2^{14}$ simple harmonic oscillators.
      }
  \label{Potts_256_SHO_16384}
  \end{subfigure}
 \begin{subfigure} [htb] {0.3\textwidth}
 \includegraphics[width=\columnwidth]{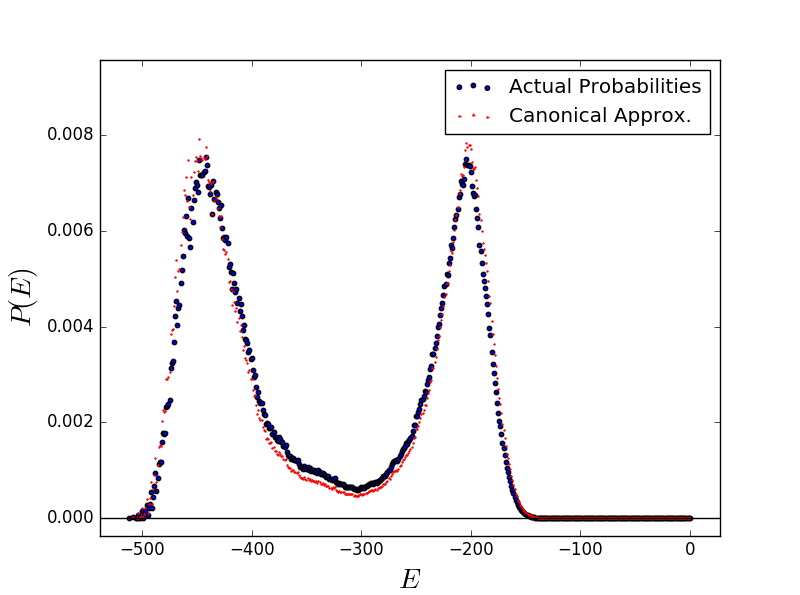}
  \caption{
 $2^{16}$ simple harmonic oscillators.
      }
  \label{Potts_256_SHO_65536}
\end{subfigure}
\end{center}
\caption{  Comparison of the canonical distribution (black dots)
  with the true joint distribution (red dots)
  for a
  $12$-states Potts model 
  on a $16 \times 16$ lattice 
  at the  energy of the phase transition
  interacting with a system of simple harmonic oscillators.}
\label{Potts_256_SHO}
\end{figure}

For a first-order transition,
the peak in the energy becomes a double peak 
with a distance between the two maxima 
that is 
proportional to the size of the system.
The double peak structure at a first-order transition
makes the analysis of this case differ for all others.
For a small environment,
the width of the distribution is dramatically limited,
so that we see only a single peak,
as shown in 
Fig.~\ref{Potts_256_SHO_4096}.
As $N_E$ increases in 
Fig.~\ref{Potts_256_SHO_16384}
and \ref{Potts_256_SHO_65536},
the full canonical distribution, 
with the double peak,
emerges.


%

\begin{table}[H]
\caption{The values of $\delta$  
for a two-dimensional, $12$-state Potts model
at $\beta_c$,  the first-order transition.
The size of the environment
(a set of simple harmonic oscillators)
is shown in the left column.
Values of $\delta$ 
for the three sizes of the Potts model investigated
($8 \times 8$, $16 \times 16$, and $32 \times 32$) 
are given in the next three columns.
}
\begin{center}
\begin{tabular}{r|r|r|r}
$N_E$ &   $8 \times 8$  & $16 \times 16$  & $32 \times 32$  \\
\hline
$2^6$  &	2.367E-1  & 	2.467E-1  & 	2.474E-1  \\
$2^8$  	&   1.485E-1  &  	1.800E-1  &  	1.770E-1   \\
$2^{10}$  & 6.106E-2  &  	1.309E-1  &  	1.280E-1    \\
$2^{12}$   & 	1.837E-2  &  	7.995E-2  &  	9.415E-2    \\
$2^{14}$   & 	4.868E-3  &  	2.590E-2  &  	6.801E-2      \\
$2^{16}$   &  1.234E-3  &  	6.567E-3  &  	3.435E-2    \\
$2^{18}$  &	3.096E-4  &  	1.641E-3  &  	9.576E-3  \\
$2^{20}$  &	7.748E-5  &  	4.100E-4  &  	2.412E-3
\end{tabular}
\end{center}
\label{table:_Potts_optimized_beta_c_delta}
\end{table}

Data for $\delta$  
is shown in Table~\ref{table:_Potts_optimized_beta_c_delta}.
The scaling of $\delta$ with $N_E^{-1}$
is satisfied.
On the other hand,
the scaling with the size of the system, $N_S$,
is not even approximately satisfied.
This is expected, because the width of the peak in the energy 
is not proportional to $N_S^{-1/2}$.

%

\section{Conclusions}

We have shown that the microcanonical ensemble,
which assumes that the system of interest is in an energy eigenstate,
does not give a correct description of an isolated system.
If the history of the system includes 
any thermal contact with another system,
it has  a probability distribution 
that is spread over many eigenstates.
With certain reasonable conditions on the sizes
of the system of interest and the environment, 
the probability distribution is that of the canonical ensemble.
This is true 
even when the ``thermal reservoir'' is 1000 times smaller than the
system of interest.

At a second-order phase transition,
the energy peak is wider than at a point away from the transition
and
the  approach to the canonical ensemble
with increased size is slower.
However,  a macroscopic system is still well described by 
the canonical distribution
when the 
environment is substantially smaller than the system of interest.

For first-order transitions,
the double-peak probability distribution requires the environment to be 
 larger than the system,
although still not infinite.

In all cases,
the thermodynamic energy is a continuous variable,
and the validity of the canonical ensemble 
does not require an infinite thermal reservoir.

\section*{Acknowledgement}

We would like to thank Lachlan Lancaster 
for many helpful discussions.
One if us (RHS)  would  like to thank 
Roberta Klatzky 
for useful comments.
This research did not receive any specific grant from funding agencies 
in the public, commercial, or
not-for-profit sectors.

\makeatletter
\renewcommand\@biblabel[1]{#1. }
\makeatother

\bibliography{Griffin_canonical}

\end{document}